\documentclass{sig-alternate-10}

\usepackage{amsmath}
\usepackage{amsfonts}
\usepackage{amssymb}
\usepackage{wrapfig}
\usepackage{color}
\usepackage{pdfsync}
\usepackage[font=footnotesize,labelfont=bf]{caption} 
\usepackage{subcaption}
\usepackage{paralist}
\usepackage{scalerel}
\usepackage{microtype}
\usepackage{multirow}

\newtheorem{conjecture}{Conjecture}
\newtheorem{definition}{Definition}
\newtheorem{theorem}{Theorem}

\newtheorem{lemma}{Lemma}

\newcommand{\BO}[1]{{ O}\left(#1\right)}

\newcommand{\BTO}[1]{\tilde{ O}\left(#1\right)}

\newcommand{\BT}[1]{{\Theta}\left(#1\right)}
\newcommand{\BOM}[1]{\Omega\left(#1\right)}
\newcommand{\SOM}[1]{\omega\left(#1\right)}

\DeclareMathOperator*{\bigboxplus}{\scalerel*{\boxplus}{\sum}}
\DeclareMathOperator*{\bigboxtimes}{\scalerel*{\boxtimes}{\sum}}

\newcommand{\eps}{\varepsilon}

\newcommand{\U}{\mathcal{U}}
\renewcommand{\P}{{P}}
\newcommand{\Q}{{Q}}
\renewcommand{\H}{\mathcal{H}}
\newcommand{\Rbb}{\mathbb{R}}

\title{
On the Complexity of Inner Product Similarity Join\titlenote{The research leading to these results has received funding from the European Research Council under the European Union's Seventh Framework Programme (FP7/2007-2013) / ERC grant agreement no.~614331.}}

\numberofauthors{4}

\author{
\alignauthor
Thomas D. Ahle\\
\affaddr{IT University of Copenhagen}\\
\email{thdy@itu.dk}
\alignauthor
Rasmus Pagh\\
\affaddr{IT University of Copenhagen}\\
\email{pagh@itu.dk}
\and
\alignauthor
Ilya Razenshteyn\\
\affaddr{MIT CSAIL}\\
\email{ilyaraz@mit.edu}
\alignauthor
Francesco Silvestri\\
\affaddr{IT University of Copenhagen}\\
\email{fras@itu.dk}
}

\begin{document}
\maketitle
\begin{abstract}
A number of tasks in classification, information retrieval, recommendation systems, and record linkage reduce to the core problem of \emph{inner product similarity join} (IPS join): identifying pairs of vectors in a collection that have a sufficiently large inner product.
IPS join is well understood when vectors are normalized and some \emph{approximation} of inner products is allowed.
However, the general case where vectors may have any length appears much more challenging.
Recently, new upper bounds based on asymmetric locality-sensitive hashing (ALSH) and asymmetric embeddings have emerged, but little has been known on the lower bound side.
In this paper we initiate a systematic study of  inner product similarity join, showing new lower and upper bounds.
Our main results are:
\begin{itemize}
	\item Approximation hardness of IPS join in subquadratic time, assuming the strong exponential time hypothesis.
	\item New upper and lower bounds for (A)LSH-based algorithms. In particular, we show that asymmetry can be avoided by relaxing the LSH definition to only consider the collision probability of \emph{distinct} elements.
	\item A new indexing method for IPS based on linear sketches, implying that our hardness results are not far from being tight.  \end{itemize}
Our technical contributions include  new asymmetric embeddings that may be of independent interest.
At the conceptual level we strive to provide greater clarity, for example by distinguishing among signed and unsigned variants of IPS join and shedding new light on the effect of asymmetry.
\end{abstract}

\section{Introduction}

This paper is concerned with \emph{inner product similarity join} (\emph{IPS join}) where, given two sets $\P,\Q\subseteq \Rbb^d$, the task is to find for each point $q\in \Q$ at least one pair\footnote{Since our focus is on lower bounds, we do not consider the more general problem of finding all such pairs.
Also note that from an upper bound side, it is common to limit the number of occurrences of each tuple in a join result to a given number $k$.}
 $(p,q) \in \P\times \Q$ where 
the inner product (or its absolute value) is larger than a given threshold $s$.
Our results apply also to the problem where for each $q\in \Q$ we seek the vector $p\in \P$ that maximizes the inner product, a search problem known in literature as \emph{maximum inner product search} (MIPS)~\cite{RamG12,ShrivastavaL14}.

\subsubsection*{Motivation}
\emph{Similarity joins} have been widely studied in the database and information retrieval communities as a mechanism for linking noisy or incomplete data.
Considerable progress, in theory and practice, has been made to address metric spaces where the triangle inequality can be used to prune the search space (see e.g.~\cite{AndoniI08,zezula2006similarity}).
In particular, it is now known that in many cases it is possible to improve upon the quadratic time complexity of a naive algorithm that explicitly considers all pairs of tuples.
The most prominent technique used to achieve provably subquadratic running time is locality-sensitive hashing (LSH)~\cite{har2012approximate,gionis1999similarity}.
In the database community the similarity join problem was originally motivated by applications in data cleaning~\cite{Chaudhuri_ICDE06,Arasu_VLDB06}.
However, since then it has become clear that similarity join is relevant for a range of other data processing applications such as clustering, semi-supervised learning, query refinement, and collaborative filtering (see e.g.~\cite{satuluri2012bayesian} for references and further examples).
We refer to the recent book by Augsten and B{\"o}hlen~\cite{augsten2013similarity} for more background on similarity join algorithms in database systems.

\emph{Inner product} is an important measure of similarity between real vectors, particularly in information retrieval and machine learning contexts~\cite{KorenBV09,SrebroRJ05}, but not captured by techniques for metric similarity joins such as~\cite{jacox2008metric,zezula2006similarity}.
Teflioudi et al.~\cite{teflioudi2015lemp} studied the IPS join problem motivated by applications in recommender systems based on latent-factor models.
In this setting, a user and the available items are represented as vectors and the preference of a user for an item is given by the inner product of the two associated vectors.
Other examples of applications for IPS join are object detection~\cite{Felzenszwalb10} and
multi-class prediction~\cite{DeanRSSVY,JoachimsFC09}.
IPS join also captures the so-called \emph{maximum kernel search}, a general machine learning approach with applications such as image matching and finding similar protein/DNA sequences~\cite{DBLP:conf/sdm/CurtinGR13}.

\subsubsection*{Challenges of IPS join}
Large inner products do not correspond to close vectors in any metric on the vector space, so metric space techniques cannot directly be used.
In fact, there are reasons to believe that inner product similarity may be inherently more difficult than other kinds of similarity search:
Williams~\cite{Williams05,alman2015probabilistic} has shown that a truly subquadratic \emph{exact} algorithm for IPS join would contradict the Strong Exponential Time Hypothesis, an important conjecture in computational complexity.
On the upper bound side new reductions of (special cases of) \emph{approximate} IPS join to fast matrix multiplication have appeared~\cite{Valiant15,karppaSODA16}, resulting in truly subquadratic algorithms even with approximation factors asymptotically close to 1.
However, the approach of reducing to fast matrix multiplication does not seem to lead to practical algorithms, since fast matrix multiplication algorithms are currently not competitive on realistic input sizes.
From a theoretical viewpoint it is of interest to determine how far this kind of technique might take us by extending lower bounds for exact IPS join to the approximate case.

Another approach to IPS join would be to use LSH, which has shown its utility in practice.
The difficulty is that inner products do not admit locality-sensitive hashing as defined by Indyk and Motwani~\cite[Theorem 1]{ShrivastavaL14}.
Recently there has been progress on \emph{asymmetric} LSH methods for inner products, resulting in subquadratic IPS join algorithms in many settings.
The idea is to consider collisions between two \emph{different} hash functions, using one hash function for query vectors and another hash function for data vectors~\cite{ShrivastavaL14,shrivastava2015asymmetric,NeyshaburS15}.
However, existing ALSH methods give very weak guarantees in situations where inner products are small relative to the lengths of vectors.
It is therefore highly relevant to determine the possibilities and limitations of this approach.

\subsubsection*{Problem definitions}
We are interested in two variants of IPS join that slightly differ in the formulation of the objective function.
For notational simplicity, we omit the term IPS and we simply refer to IPS join as join. 
Let $s>0$ be a given value.
The first variant is the \emph{signed} join, where 
the goal is to find {at least one} pair $(p,q) \in \P\times \Q$ for each point $q\in Q$ with $p^T q\geq s$.
The second variant is the \emph{unsigned} join which finds, for each point $q\in Q$, at least one pair $(p,q) \in \P\times \Q$ where $|p^T q|\geq s$.
We observe that the unsigned version can be solved with the signed one by computing the join between $\P$ and $\Q$ and between $\P$ and $-\Q$, and then returning only pairs where the absolute  inner products are larger than $s$.
Signed  join is of interest when searching for similar or preferred items with a positive correlation, like in recommender systems.
On the other hand, unsigned join can be used when studying relations among phenomena where even a large negative correlation is of interest.
We note that previous works do not make the distinction between the signed and unsigned versions since they focus on settings where there are no negative dot products.

Our focus is on \emph{approximate} algorithms for signed and unsigned joins. 
Indeed, approximate algorithms allow us to overcome, at least in some cases, the curse of dimensionality without significantly affecting the final results.
Approximate signed joins are defined as follows.
\begin{definition}[Approximate signed join]
Given two point sets $\P$, $\Q$ and values $0<c<1$ and $s>0$, the \emph{signed $(cs,s)$ join} returns, for each $q\in Q$, at least one pair $(p,q)\in\P\times\Q$ with $p^Tq\geq cs$ if there exists $p'\in P$ such that $p'^Tq\geq s$. No guarantee is provided for $q\in Q$ where there is no $p'\in P$ with $p'^T q\geq s$.
\end{definition}
The unsigned $(cs,s)$ join is defined analogously by taking the absolute value of dot products.
Indexing versions of signed/unsigned exact/approximate joins can be defined in a similar way.
For example, the signed  $(cs,s)$ search is defined as follows: given a set $\P\subset \mathbb{R}^d$ of $n$ vectors, construct a data structure that efficiently returns a vector $p\in \P$ such that $p^T q> cs$  for any given query vector $q\in \mathbb{R}^d$, under the promise that there is a point $p'\in P$ such that $p^Tq\geq s$ (a similar definition holds for the unsigned case).

As already mentioned,  LSH is often used for solving similarity joins.  
In this paper, we use the following definition of asymmetric LSH based on the definition in~\cite{ShrivastavaL14}.
\begin{definition}[Asymmetric LSH]
Let $\U_p$ denote the data domain and $\U_q$ the query domain. Consider a family $\H$ of pairs of hash functions $h=(h_p(\cdot), h_q(\cdot))$. Then $\H$ is said \emph{$(s,cs,P_1,P_2)$-asymmetric LSH} for a similarity function $sim$ if for any $p\in \U_p$ and $q\in \U_q$ we have: 
\begin{enumerate}
\item if $sim(p,q)\geq s$ then $\Pr_{\H}[h_p(p)=h_q(q)]\geq P_1$; 
\item if $sim(p,q)< cs$ then $\Pr_{\H}[h_p(p)=h_q(q)]\leq P_2$.
\end{enumerate}
\end{definition}

When $h_p(\cdot)=h_q(\cdot)$, we get the traditional (symmetric) LSH definition.
The $\rho$ value of an (asymmetric) LSH is defined as usual with $\rho=\log P_1/\log P_2$~\cite{AndoniI08}.
Two vectors $p\in \U_p$ and $q\in \U_q$ are said to collide under a hash function from  $\H$ if $h_p(p)=h_q(q)$.

\subsection{Overview of results}

\subsubsection*{Hardness results}
The first results of the paper are conditional lower bounds for approximate signed and unsigned IPS join that rely on a conjecture about the \emph{Orthogonal Vectors Problem} (OVP).
This problem consists in determining if two sets $A,B \subseteq \{0,1\}^d$, each one with $n$ vectors, contain $x\in A$ and $y\in B$ such that $x^Ty=0$.
It is  conjectured that there cannot exist an algorithm that solves OVP in $\BO{n^{2-\epsilon}}$ time as soon as $d=\SOM{\log n}$, for any given constant $\epsilon >0$. 
Indeed, such an algorithm would imply that the Strong Exponential Time Hypothesis (SETH) is true~\cite{Williams05}.

Many recent interesting hardness results rely on reductions from OVP, however we believe ours is the first example of using the conjecture to show the conditional hardness for an approximate problem.
In particular we show the following result:
\begin{theorem}\label{thm:nojoin}
Let $\alpha > 0$ be given and consider sets of vectors $P,Q$ with $|Q|=n$, $|P|=n^\alpha$.
Suppose there exists a constant $\epsilon>0$ and an algorithm with running time at most $d^{\BO{1}}n^{1+\alpha-\epsilon}$, when $d$ and $n$ are sufficiently large and for all $s>0$, for at least one of the following IPS join problems:
\begin{enumerate}
\item Signed $(cs,s)$ join of $P,Q \subseteq \{-1,1\}^d$ with $c > 0$.
\item Unsigned $(cs,s)$ join of $P,Q \subseteq \{-1,1\}^d$ with $c=e^{-o(\sqrt{\log n}/\log\log n)}$.
\item Unsigned $(cs,s)$ join of $P,Q \subseteq \{0,1\}^d$ with $c=1-o(1)$.
\end{enumerate}
Then the OVP conjecture is false.
\end{theorem}

{\bf Discussion.}

For the search problem, theorem~\ref{thm:nojoin} implies that, assuming the OVP conjecture, there does not exist a data structure for signed/unsigned $(cs,s)$ inner product search with $(nd)^{\BO{1}}$ construction time and $n^{1-\epsilon}d^{\BO{1}}$ query time, for constant $\epsilon>0$.
This follows by considering a join instance with $\alpha$ constant small enough that we can build the data structure on $\P$ in time $o(nd)$.
We can then query over all the points of $\Q$ in time $n^{1+\alpha(1-\epsilon)}d^{\BO{1}}$, contradicting the OVP conjecture.
Our result can be seen as an explanation of why all LSH schemes for IPS have failed to provide sub-linear query times for small~$s$.
As the theorem however does not cover the case where $c$ is very small, e.g. $n^{-\delta}$ for unsigned $\{-1,1\}^d$, we show in section 4 that for such approximation requirements, there are indeed useful data structures.

We stress that the hardness result holds for algorithms solving signed/unsigned $(cs,s)$ joins for \emph{any} $c$ in the specified range and \emph{all} $s>0$.
It is possible to show complete relations between hard values of $c$, $s$ and $d$, but for the sake of clearnes, we have prefered to optimize the largest range of hard $c$'s.
For intuition we can say, that the exact instances of $(cs,s)$ joins that are hard, turn out to be the ones where $s/d$ and $cs/d$ are very small,
that is when we have to distinguish nearly orthogonal vectors from very nearly orthogonal vectors.
If we inspect the proofs of Theorem~\ref{thm:nojoin} and Lemma~\ref{lem:embeddings}, we see that for unsigned join in $\{-1,1\}^d$, the hard case has $cs/d$ around $n^{1/\log\log n}$.
Similarly for $\{0,1\}^d$ join, $cs$ ends up at just barely $\omega(1)$, while the $d$ is as high as $n^{o(1)}$.
It is interesting to note for $\{0,1\}$ that if $cs$ had been slightly lower, at $O(1)$, we could have solved the OVP problem exact in subquadratic time using an $n{n^{o(1)}\choose O(1)}=n^{1+o(1)}$ algorithm.

It is interesting to compare our conditional lower bound to the recent upper bounds by Karppa et al.~\cite{karppaSODA16}, who get sub-quadratic running time for unsigned join of normalized vectors in $\{-1,1\}^d$, when $\log(s/d)/\log(cs/d)$ is a constant smaller than 1.\footnote{More precisely they need $\log(s/d)/\log(cs/d)<2/\omega$, where $\omega$ is the matrix multiplication constant. Note that the $d$ term is due to normalization.}
Our next Theorem~\ref{thm:nojoin2} shows that we cannot hope to do much better than this, though it does not completely close the gap.
A hardness result for $\log(s/d)/\log(cs/d)=1-o(1)$ is still an interesting open problem.
However, while the algorithm of Karppa et al. works even for dimension $n^{1/3}$, our bound only requires the dimension to be slightly larger than polylog, so it may well be that their algorithm is optimal, while another algorithm with a higher dependency on the dimension matches our bound from above.

\begin{theorem}\label{thm:nojoin2}
Let $\alpha > 0$ be given and consider sets of vectors $P,Q$ with $|Q|=n$, $|P|=n^\alpha$.
Suppose there exists a constant $\epsilon>0$ and an algorithm with running time at most $d^{\BO{1}}n^{1+\alpha-\epsilon}$, when $d$ and $n$ are sufficiently large and for all $s>0$, for at least one of the following IPS join problems:
   \begin{enumerate}
   \item Unsigned $(cs,s)$ over $\{-1,1\}^d$
      where $\frac{\log(s/d)}{\log(cs/d)}=1-o(1/\sqrt{\log n})$
   \item Unsigned $(cs,s)$ over $\{0,1\}^d$
      where $\frac{\log(s/d)}{\log(cs/d)}=1-o(1/\log n)$
   \end{enumerate}
   Then the OVP conjecture is false.
\end{theorem}

The $\{-1,1\}^d$ case seems to be harder than the $\{0,1\}^d$ case.
In fact Valiant~\cite{Valiant15} reduces the general case of $P,Q\subseteq\mathbb{R}^d$ to the case $P,Q\subseteq \{-1,1\}^d$ using the Charikar hyperplane LSH~\cite{charikar2002similarity}.
Another piece of evidence is that we can achive runtime $n^{1+\frac{\log(s/d)}{\log(cs/d)}}$ using LSH for $\{0,1\}^d$, but it is not known to be possible for $\{-1,1\}^d$.
Furthermore there appears to be some hope for even better data dependent LSH, as we show in section~\ref{up2}.
The $\{0,1\}^d$ case is particularly interesting, as it is occurs often in practice, for example when the vectors represent sets.
A better understanding of the upper and lower bounds for this case is a nice open problem.
For an ellaborate comparison of the different upper and lower bounds, see Table~\ref{tab:bounds}.

{\bf Techniques.}
From a technical point of view, the proof uses a number of different algebraic techniques to expand the gap between orthogonal and non-orthogonal vectors from the OVP problem.
For the $\{-1,1\}$ we use an enhanced, deterministic version of the ``Chebyshev embedding''~\cite{Valiant15}, while for the interesting $\{0,1\}$ part, we initiate a study of  embeddings for restricted alphabets.

\begin{table*}[t]
{\footnotesize
   \begin{tabular}{l|ll|ll}
   \hline
      {\rule[-.3cm]{0cm}{.8cm}}
      Problem
      & Hard approx.
      & Permissible approx.
      & Hard approx.
      & Permissible approx.\\
   \hline
      {\rule[-.3cm]{0cm}{.8cm}}
      Signed $(cs,s)$ over $\{-1,1\}^d$
      & $c > 0$
      & -
      & $\frac{\log(s/d)}{\log(cs/d)} > 0$
      & -
      \\
   \hline
      {\rule[-.3cm]{0cm}{.8cm}}
      Unsigned $(cs,s)$ over $\{-1,1\}^d$
      & $c\ge e^{-o(\frac{\sqrt{\log n}}{\log\log n})}$
      & $c < n^{-\epsilon}$~\cite{karppaSODA16}
      & $\frac{\log (s/d)}{\log (cs/d)} \ge 1-o(\frac1{\log n})$ \cite{karppaSODA16}
      & $\frac{\log (s/d)}{\log (cs/d)} = 1-\epsilon$ \cite{karppaSODA16}
      \\
      {\rule[-.3cm]{0cm}{.8cm}}
      &
      & $c < n^{-\epsilon}$
      & $\frac{\log (s/d)}{\log (cs/d)} \ge 1-o(\frac1{\sqrt{\log n}})$
      & $\frac{\log s/d}{\log cs/d} = 1/2 - \epsilon$
      \\
   \hline
      {\rule[-.3cm]{0cm}{.8cm}}
      Unsigned $(cs,s)$ over $\{0,1\}^d$
      & $c\ge1-o(1)$
      & $c < n^{-\epsilon}$
      & $\frac{\log s/d}{\log cs/d}\ge1-o(\frac{1}{\log n})$
      & $\frac{\log s/d}{\log cs/d} = 1-\epsilon$
      \\
   \hline
   \end{tabular}
   }
   \caption{
      The table describes the ranges of approximations that are hard, when parametrized in terms of $c$ (second and third column) or $\log(s/d)/\log(cs/d)$ ratio (fourth and fifth column).
      Any algorithm for subquadratic join, which overlap with these ranges, would contradict the OVP.
      The permissible approximations are those ranges for which truly subquadratic algorithms are known.
      The upper bounds cited to \cite{karppaSODA16} use fast matrix multiplication, whereas the rest don't and are usable as data structures.
      The bounds not cited elsewhere are new in this paper, though we are aware that other people have noted the hardness of signed $\{-1,1\}$ join and the data structure for $\{0,1\}$ join.\label{tab:bounds}
      \label{tab:bounds}}
\end{table*}

\subsubsection*{Inner product LSH lower bounds}
In the second part of the paper we focus on LSH functions for signed and unsigned IPS. 
We investigate the gap between the collision probability $P_1$ of vectors with inner product (or absolute inner product) larger than $s$ and the collision probability $P_2$ of vectors with inner product (or absolute inner product) smaller than $cs$. 
As a special case, we get the impossibility result in~\cite{NeyshaburS15,ShrivastavaL14}, that there cannot exist an asymmetric LSH for unbounded query vectors.
Specifically we get the following theorem:

\begin{theorem}\label{th:gap}
   Consider an $(s,cs,P_1,P_2)$-asymmetric LSH for signed IPS when data and query domains are  $d$-dimensional balls with unit radius and radius $U$ respectively.
   Then, the following upper bounds on $P_1-P_2$ apply:
   \begin{enumerate}
      \item if $d\geq 1$ and $s\leq \min\{cU, U/(4\sqrt{d}\}$, we have  $P_1-P_2 = \BO{{1}/{\log (d\log_{1/c} (U/s))}}$ for signed and unsigned IPS; 
      \item if $d\geq 2$ and $s\leq U/(2d)$, we have $P_1-P_2=\quad$ $\BO{{1}/{\log (dU/(s(1-c)))}}$ for signed IPS;
      \item if $d > \BT{U^5/(c^2s^5)}$ and $s\leq U/8$,  we have $P_1-P_2 = \BO{{\sqrt{s/U}}}$ for signed and unsigned IPS.
   \end{enumerate}

   It follows that, for any given dimension $d$,  there cannot exist an asymmetric LSH when the query domain is unbounded.
\end{theorem}

{\bf Discussion.}
The upper bounds for  $P_1-P_2$ translate into lower bounds for the $\rho$ factor, as soon as $P_2$ is fixed. 
To the best of our knowledge, this is the first lower bound on $\rho$ that holds for asymmetric LSH. 
Indeed, previous results~\cite{MotwaniNP06,ODonnellWZ14} have investigated lower bounds for symmetric LSH and it is not clear if they can be extended to the asymmetric case. 

{\bf Techniques.}
The starting point of our proof is the same as in~\cite{NeyshaburS15}: Use a collision matrix given by two sequences of data and query vectors that force the gap to be small.
The proof in~\cite{NeyshaburS15} then applies an asymptotic analysis of the margin complexity of this matrix~\cite{SrebroS05}, and it shows that  for any given value of $P_1-P_2$ there are sufficiently large data and query domains for which the gap must be smaller. 
Unfortunately, due to their analysis, an upper bound on the gap for a given radius $U$ of the query domain is not possible, and so the result does not rule out very large gaps for small domains. 
Our method also highlights a dependency of the gap on the dimension, which is missing in~\cite{NeyshaburS15}. 
In addition, our proof 
holds for $d=1$ and only uses purely combinatorial arguments.

\subsubsection*{IPS upper bounds}

In the third part we provide some insights on the upper bound side.
We first show that it is possible to improve the asymmetric LSH in~\cite{NeyshaburS15,shrivastava2015asymmetric} by just plugging the best known data structure for Approximate Near Neighbor for $\ell_2$ on a sphere~\cite{andoni2015optimal} into the reduction in~\cite{NeyshaburS15,Bachrach14}. With data/query points in the unit ball, this LSH reaches $\rho=(1-s)/(1+(1-2c)s)$.
In the $\{0,1\}$ domain, this LSH improves upon the state of the art \cite{shrivastava2015asymmetric} for some ranges of $c$ and $s$.

Then we show how to circumvent the impossibility results in~\cite{NeyshaburS15,ShrivastavaL14} by showing that there exists a symmetric LSH when the data and query space coincide by allowing the bounds on collision probability to not hold when the  data and query vectors are identical.

We conclude by describing a data structure based on the linear sketches for $\ell_p$ in~\cite{alexlpmanuscript} for unsigned $(cs,s)$ search: for any given $0 < \kappa \leq 1/2$, the data structure yields a $c=1/n^{\kappa}$ approximation with $\BTO{dn^{2-2/\kappa}}$ construction time and $\BTO{dn^{1-2/\kappa}}$ query time.
Theorem~\ref{thm:nojoin} suggests that we cannot substantially improve the approximation with similar performance.

The last data structure allows us to reach truly subquadratic time for $c=1/n^{\kappa}$ for the unsigned version in the $\{0,1\}$ and $\{-1,1\}$ domains for all value $s$. 
We note that the result in \cite{karppaSODA16} also reaches subquadtratic time for the $\{-1,1\}$ case.
However, it exploits fast matrix multiplication, whereas our data structure does not.


\subsection{Previous work}

\subsubsection*{Similarity join}
Similarity join problems have been extensively studied in the database literature (e.g.~\cite{Chaudhuri_ICDE06,chen2007efficient,Cohen_TKDE01,jacox2008metric,jiang2013efficient,li2011pass,lu2013string,silva2010similarity,wang2011fast,wang2012can,xia2004gorder}), as well as in information retrieval (e.g.~\cite{Bayardo_WWW07,Das_WWW07,Xiao_WWW08}), 
and knowledge discovery (e.g.~\cite{achlioptas2011two,Bahmani_CIKM12,Wang_KDD13,zadeh2013dimension,zhang2008fastanova}).
Most of the literature considers algorithms for particular metrics (where the task is to join tuples that are near according to the metric), or particular application areas (e.g.
near-duplicate detection).
A distinction is made between methods that approximate distances in the sense that we only care about distances up to some factor $c > 1$, and methods that consider exact distances.
Known exact methods do not guarantee subquadratic running time.
It was recently shown how approximate LSH-based similarity join can be made I/O-efficient~\cite{PaghPSS15}.

\subsubsection*{IPS join}
The inner product similarity for the case of \emph{normalized} vectors is known as ``cosine similarity'' and it is well understood~\cite{Cha02,lee2010efficient,RamG12}.
While the general case where vectors may have any length appears theoretically challenging,
\emph{practically} efficient indexes for unsigned search were proposed in~\cite{RamG12,KoenigsteinRS12}, based on tree data structures combined with a branch-and-bound space partitioning technique similar to $k$-d trees, and in~\cite{Bachrach14} based on principal component axes trees.
For document term vectors Low and Zheng~\cite{low2012fast} showed that unsigned search can be sped up using matrix compression ideas.
However, as many similarity search problems, the exact version considered in these papers suffers from the curse of dimensionality~\cite{WeberSB98}.

The efficiency of approximate IPS approaches based on LSH is studied in~\cite{ShrivastavaL14,NeyshaburS15}.
These papers show that a traditional LSH does exist when the data domain is the unit ball and the query domain is the unit sphere, while it does not exist when both domains are the unit ball (the claim automatically applies to any radius by suitably normalizing vectors).
On the other hand an asymmetric LSH exists in this case, but it cannot be extended to the unbounded domain $\mathbb{R}^d$.
An asymmetric LSH for binary inner product is proposed in~\cite{shrivastava2015asymmetric}.
The unsigned version is equivalent to the signed one when the vectors are non-negative.


\subsubsection*{Algebraic techniques}
Finally, recent breakthroughs have been made on the (unsigned) join problem in the approximate case as well as the exact.
Valiant~\cite{Valiant15} showed how to reduce the problem to matrix multiplication, when $cs\approx O(\sqrt{n})$ and $s\approx O(n)$, significantly improving on the asymptotic time complexity of approaches based on LSH.
Recently this technique was improved by Karppa et al.~\cite{karppaSODA16}, who also generalized the sub-quadratic running time to the case when $\log(s)/\log(cs)$ is small.
In another surprising development Alman and Williams~\cite{alman2015probabilistic} showed that for $d=\BO{\log n}$ dimensions, truly subquadratic algorithms for the \emph{exact} IPS join problem on binary vectors is possible.
Their algorithm is based on an algebraic technique (probabilistic polynomials) and tools from circuit complexity.

\section{Hardness of IPS join}
We first provide an overview of OVP and of the associated conjecture in next Section~\ref{sec:prelOVP}. Then, in Section~\ref{sec:red}, we prove Theorem~\ref{thm:nojoin} by describing some reductions from the OVP to signed/unsigned joins.

\subsection{Preliminaries}\label{sec:prelOVP}
The Orthogonal Vectors Problem (OVP) is defined as follows:
\begin{definition}[OVP]
Given two sets $P$ and $Q$, each one containing $n$ vectors in $\{0,1\}^d$, detect if there exist vectors $p \in P$ and $q \in Q$ such that $p^Tq = 0$.
\end{definition}
OVP derives its hardness from the Strong Exponential Time Hypothesis (Williams~\cite{Williams05}), but could potentially be true even if SETH is not.
 We will therefore assume the following plausible conjecture:\footnote{
    We will use the name, OVP, for the problem as well as the conjecture.
Sorry about that.}
\begin{conjecture}[OVP, \cite{Williams05}]
 \label{conj:ovp}
 For every constant $\epsilon > 0$, there is no algorithm for OVP with $|P|=|Q|=n$ and dimension $d=\omega(\log n)$ running in $O(n^{2-\epsilon})$ time.
\end{conjecture}
The conjecture does not hold for $d=O(\log n)$: recently Abboud et al.~\cite{AbboudWY15} have proposed an algorithm for OVP running in time $n^{2-1/O(\gamma\log^2\gamma)}$, when $d=\gamma\log n$.
Thus, in order to disprove OVP, an algorithm must be strongly subquadratic when $d=\gamma\log n$ for \emph{all} constant $\gamma>0$.

The OVP conjecture, as usually stated, concerns the case where the two sets have equal size.
However in order to eventually show hardness for data structures, we consider the following generalization of OVP, which follows directly from the original:
\begin{lemma}[Generalized OVP]\label{lem:genovp}
 Suppose that there exist constants $\epsilon>0$ and $\alpha>0$, and an algorithm such that for $d=\SOM{\log n}$ the algorithm solves OVP for $P,Q\subseteq \{0,1\}^{d}$ where $|P|=n^\alpha$ and $|Q|=n$ in time $O(n^{1+\alpha-\epsilon})$.
 Then OVP is false.
\end{lemma}
\begin{proof}
 Without loss of generality assume $\alpha \leq 1$ (otherwise is enough to invert the role of $P$ and $Q$).
 Suppose we have an algorithm running in time $O(n^{1+\alpha-\epsilon})$ for some $\epsilon>0$.
 Take a normal OVP instance with $|P|=|Q|=n$.
 Split $P$ into chunks $P_i$ of size $n^{\alpha}$ and run the OVP algorithm on all pairs $(P_i,Q)$.
 By our assumption this takes time $n^{1-\alpha} O(n^{1+\alpha-\epsilon}) = O(n^{2-\epsilon})$, contradicting OVP.
\end{proof}
\subsection{Reductions from OVP}\label{sec:red}

In this section we prove Theorem~\ref{thm:nojoin}, about hardness of approximate joins.
We will do this by showing the existence of certain efficient `gap embeddings' that make orthogonality discoverable with joins.
We need the following definition:
\begin{definition}[Gap Embedding]
 An unsigned ($d_1$, $d_2$, $cs$, $s$)-gap embedding into the domain $\mathcal{A}$ is a pair of functions $(f,g) : \{0,1\}^{d_1} \to \mathcal{A}^{d'_2}$, where $d'_2\le d_2$, $\mathcal A\subseteq\mathbb R$, and for any $x,y\in\{0,1\}^{d_1}$:
 \begin{align*}
 | f(x)^T g(y)| &\ge s
 \quad\text{when}\quad  x^Ty = 0\\
 | f(x)^T g(y)| &\le cs
 \quad\text{when}\quad x^T y \ge 1
 \end{align*}
 A `signed embedding' is analogous, but without the absolute value symbols.
 We further require that the functions $f$ and $g$ can be evaluated in time polynomial to $d_2$.
\end{definition}
Gap embeddings connect to the join problem, by the following technical lemma: 
\begin{lemma}
 \label{lem:emb+join}
 Suppose there exist a join algorithm for (un)signed $(cs,s)$-join over $\mathcal{A}$ and a family of (un)signed $(d, 2^{o(d)}, cs, s)$-gap embeddings into $\mathcal{A}$, for all $d$ large enough.

 \begin{itemize}
 \item For given constants $\alpha\geq 0$, and $\epsilon> 0$, the algorithm has running time $d^{O(1)}n^{1+\alpha-\epsilon}$ when $|Q|=n$ and $|P|=n^\alpha$ for all $n$ and $d$ large enough.
 \item The embedding has can be evaluated in time $d_2^{O(1)}$.
 \end{itemize}
Then OVP can be solved in $n^{1+\alpha-\epsilon}$ time, and the conjecture is false.
\end{lemma}
\begin{proof}
   First notice that for any function $d_2=2^{o(d)}$ we can take $d=\omega(\log n)$ growing slowly enough that $d_2=n^{o(1)}$.

   To see this, assume $d_2(d)=2^{f(d)}$ where $f(d)=o(d)$.
   Then we have $f(d(n))=o(d(n))=o(1)d(n)$ that is $f(d(n))=f'(n)d(n)$ for some $f'(n)=o(1)$.
   Now take $d(n)=\frac{\log n}{\sqrt{f'(n)}}=\omega(\log n)$ and we get $d_2(d(n))=2^{f'(n)d(n)}=2^{\sqrt{f'(n)}\log n}=n^{o(1)}$ as desired.

   Hence, there is a family of $(d(n), d_2(n), cs, s)$-gap embeddings for all $n$, where $d(n)=\omega(\log n)$ and $d_2(n)=n^{o(1)}$.
   By the generalized OVP lemma, for large enough $n$, we can thus take a hard OVP instance with $|Q|=n$, $|P|=n^\alpha$ and dimension $d(n)$.
   Apply the coresponding gap embedding, $(f,g)$, to the instance, such that the maximum inner product between $f(P)$, $g(Q)$ is at least $s$ if the OVP instance has an orthogonal pair and $\le cs$ otherwise.
 Now run the algorithm for (un)signed $(cs,s)$ join on $(f(P),g(Q))$, which produces the orthogonal pair, if it exists.

 It remains to show that the running time of the above procedure is $O(n^{1+\alpha-\epsilon'})$ for some $\epsilon'>0$.
 But this is easy, since by assumption, performing the embedding takes time $n^{1+o(1)}$,
 and nunning the algorithm on vectors of dimension $n^{o(1)}$ takes time $n^{1+\alpha-\epsilon+o(1)}$.
 So letting $\epsilon'=\epsilon/2$ suffices.
\end{proof}

The last ingredient we need to show Theorem~\ref{thm:nojoin} is a suitable family of embeddings to use with Lemma~\ref{lem:emb+join}:

\begin{lemma}\label{lem:embeddings}
 We can construct the following gap embeddings:
 \begin{enumerate}
 \item A signed $(d, 4d-4, 0, 4)$-embedding into $\{-1,1\}$.
 \item An unsigned $(d, (9d)^q, (2d)^q, (2d)^q e^{q/\sqrt{d}}/2)$-embedding into $\{-1,1\}$, for any $q\in\mathbb{N_+}$, $d>1$.
 \item An unsigned $(d, k2^{d/k}, k-1, k)$-embedding into $\{0,1\}$, for any integer $1\le k\le d$.
 \end{enumerate}
\end{lemma}

\begin{proof}
 We will use the following notation in our constructions:
 Let $x\boxplus y$ be the concatenation of vectors $x$ and $y$;\footnote{
 $\boxplus$ for concatenation and $\boxtimes$ for tensoring stresses their dual relationship with $+$ and $\times$ on the inner products in the embedded space.
 We note however that in general, it is only safe to commute $\boxplus$'es and $\boxtimes$'es in an embedding $(f,g)$, when both $f$ and $g$ are commuted equally.}
 Let $x^n$ mean $x$ concatenated with itself $n$ times;\footnote{
 If we wanted to further stress the duality between construction and embedding, we could define $\vec{n}$ to be the all $1$ vector of length $n$.
 Then $\vec{n}\boxtimes x$ would stand for repeating $x$ $n$ times.}
 And let $x\boxtimes y$ mean the vectorial representation of the outer product $xy^T$.
 Tensoring is interesting because of the following folklore property:
 $(x_1\boxtimes x_2)^T(y_1\boxtimes y_2)
 = \text{trace }(x_1x_2^T)^T(y_1y_2^T)
 = \text{trace }x_2(x_1^Ty_1)y_2^T
 = (x_1^Ty_1)(x_2^Ty_2)$.

 \item\emph{(Embedding 1)}
 The signed embedding is a simple coordinate wise construction:
 \begin{align*}
 \hat{f}(0) &:= (\phantom{-}1,-1,-1) &
 \hat{g}(0) &:= (\phantom{-}1,\phantom{-}1,-1)\\
 \hat{f}(1) &:= (\phantom{-}1,\phantom{-}1,\phantom{-}1) &
 \hat{g}(1) &:= (-1,-1,-1)
 \end{align*}
 such that $\hat{f}(1)^T\hat{g}(1)=-3$ and $\hat{f}(0)^T\hat{g}(1)=\hat{f}(1)^T\hat{g}(0)=\hat{f}(0)^T\hat{g}(0)=1$.
 This, on its own, gives a $(d, 3d, d-4, d)$ embedding, as non orthogonal vectors need to have at least one (1,1) at some position.

 We can then translate all the inner products by $-(d-4)$:
 \begin{align*}
 f(x) &:= \hat{f}(x_1) \boxplus \dots \boxplus \hat{f}({x_n}) \boxplus 1^{d-4} \\
 g(x) &:= \hat{g}(x_1) \boxplus \dots \boxplus \hat{g}({x_n}) \boxplus (-1)^{d-4}
 \end{align*}
 which gives the $(d,4d-4,0,4)$ embedding we wanted.
 Note that the magnitudes of non orthogonal vectors may be large ($-4d+4$), but we do not care about those for signed embeddings.

 \item\emph{(Embedding 2)}
 We recall the recursive definition of the $q$-th order Chebyshev polynomial of first kind, with $q\geq 0$ (see, e.g., \cite{abramowitz1964handbook} page 782):
 \begin{align*}
 T_0(x) & = 1 \\
 T_1(x) & = x \\
 T_{q}(x) & = 2xT_{q-1}(x) - T_{q-2}(x)
 \end{align*}
 The polynomials have the following properties \cite{Valiant15}:
 \begin{align*}
 |T_q(x)| &\le 1 \quad\text{when}\quad |x|\le 1\\
 |T_q(1+\epsilon)| &\ge e^{q \sqrt{\epsilon}} \quad\text{when}\quad 0<\epsilon<1/2
 \end{align*}

 We use the same coordinate wise transformation as in the signed embedding, but instead of translating by a negative value, we translate by adding $d+2$ ones, giving a $(d,4d+2,2d-2,2d+2)$ unsigned embedding.
 Let the vectors created this way be called $x$ and $y$.

 On top of that, we would like to construct an embedding for the polynomial $T_q(u/2d)$, where $T_q$ is the $q$th order Chebyshev polynomial of the first kind.
 However since this will not in general be interger, there is no hope for constructing it using $\{-1,1\}$.

 Luckily it turns out we can construct an embedding for $b^qT_q(u/b)$ for any integers $b$ and $q$.
 Let $(f_q,g_q)$ be the $q$th embedding of this type, defined by:
 \begin{align*}
 & f_0(x),\ g_0(y) := 1,\ 1\\
 & f_1(x),\ g_1(y) := x,\ y\\
 & f_{q}(x) := (x\boxtimes f_{q-1}(x))^2\boxplus f_{q-2}(x)^{(2d)^2}\\
 & g_{q}(y) := (y\boxtimes g_{q-1}(y))^2\boxplus (-g_{q-2}(y))^{(2d)^2}
 \end{align*}

 We make the following observations:
\begin{itemize}
 \item If $x$ and $y$ are $\{-1,1\}$ vectors, then so are $f_q(x)$ and $g_q(y)$.

 \item The inner product of the embedded vectors, $f_q(x)^Tg_q(x)$ is a function of the original inner product:
 \begin{align*}
 f_0(x)^Tg_0(y) &= 1\\
 f_1(x)^Tg_1(y) &= x^Ty\\
 f_{q}(x)^Tg_{q}(y) &= 2x^Ty\ f_{q-1}(x)^Tg_{q-1}(y) \\&\quad-(2d)^2f_{q-2}(x)^Tg_{q-2}(y)
 \end{align*}

 Indeed it may be verified from the recursive definition of $T_q$ that $f_q(x)^Tg_q(y) = (2d)^nT_q(x^Ty/2d)$ as wanted.

 \item Let $d_q$ be the dimension of $f_q(x)$ and $g_q(y)$.
 Then we have:
 \begin{align*}
 d_0 &= 1\\
 d_1 &= 4d-4\\
 d_q &= 2(4d-4) d_{q-1} + (2d)^2 d_{q-2}
 \end{align*}
 It can be verified that $d_q\le(9d)^q$ for any $q\ge0$ and $d\ge8$.
 Interestingly the $(2d)^2$ concatenations don't increase $d_q$ significantly, while $d^{2+\epsilon}$ for any $\epsilon>0$ would have killed the simple exponential dependency.

 \item Finally, with dynamic programming, we can compute the embeddings in linear time in the output dimension.
 This follows from induction over $q$.
\end{itemize}

 Putting the above observations together, we have for any integer $q\ge0$ a $(d, (9d)^q, (2d)^q, (2d)^qT_q(1+1/d))$ embedding.
 By the aforementioned properties of the Chebyshev polynomials, we have the desired embedding.
 We note that the Chebyshev embedding proposed by Valiant~\cite{Valiant15} can provide similar results; however, our construction is deterministic, while Valiant's is randomized.

 \item\emph{(Embedding 3)}
 The third embedding maps into $\{0,1\}$.
	The difficulty here is that without $-1$, we cannot express subtraction as in the previous argument.
 It turns out however, that we can construct the following polynomial:
 \begin{align*}
 (1-x_1y_1)(1-x_2y_2)\cdots(1-x_dy_d)
 \end{align*}
 since $\{0,1\}$ is closed under tensoring and
 \begin{align*}
 1-x_i y_i=( 1-x_i, 1)^T( y_i, 1-y_i)
 \end{align*}
 where $1-x_i$, $y_i$ and $1-y_i$ are both in $\{0,1\}$.
 The polynomial has the property of being $1$ exactly when the two vectors are orthogonal and $0$ otherwise.

 However we cannot use it directly with Lemma \ref{lem:emb+join}, as it blows up the dimension too much, $d_2=2^{d_1}$.
 Instead we ``chop up'' the polynomial in $k$ chunks and take their sum:
 \begin{align*}
 \sum_{i=0}^{k-1}\prod_{j=1}^{d/k}(1-x_{ik/d+j}y_{ik/d+j})
 \end{align*}
 This uses just $d_2=k2^{d/k}$ dimensions, which is more manageble.
 If $k$ does not divide $d$, we can let the last ``chop'' of the polynomial be shorter than $d/k$, which only has the effect of making the output dimension slightly smaller.

 Finally we get the gap $s=k$ and $cs= k-1$.
 The later follows because for non orthogonal vectors, at least one chunk has a $(1-x_iy_i)$ terms which evaluates to zero.
 We thus have a $(d, k2^{d/k}, k-1, k)$-embedding into $\{0,1\}$.
 The explicit construction is thus:
 \begin{align*}
 f(x) &:= \bigboxplus_{i=0}^{k-1}\bigboxtimes_{j=1}^{d/k}(1-x_{ik/d+j}, 1)\\
 g(x) &:= \bigboxplus_{i=0}^{k-1}\bigboxtimes_{j=1}^{d/k}( y_{ik/d+j}, 1-y_{ik/d+j})
 \end{align*}
 And the running time is linear in the output dimension.
\end{proof}

Finally we parametize and prove Theorem~\ref{thm:nojoin}.
\begin{proof} (Theorem~\ref{thm:nojoin})
   We first prove the bounds parametrized by $c$.
   To get the strongest possible results, we want to get $c$ as small as possible, while keeping the dimension bounded by $n^\delta$ for some $\delta>0$.
   \begin{enumerate}
      \item The first embedding is already on the form $(d, 2^{o(d)}, 0, 4)$, showing theorem 1 for signed $(0,4)$ join, and thus any $c>0$.

   \item In the second embedding we can take $q$ to be any function in $o\left(\frac{d}{\log d}\right)$, giving us a family of ($d$, $2^{o(d)}$, $2^{o(d)}$, $2^{o(d)}e^{o\left(\frac{\sqrt{d}}{\log d}\right)}$).
      Thus by lemma~\ref{lem:emb+join}, and for a small enough $d=\omega(\log n)$ in OVP, we get that even $c\le e^{-o\left(\frac{\sqrt{\log n}}{\log\log n}\right)} \le 1/\text{polylog(n)}$ is hard.

      \item Finally for the third embedding, we can pick any $k=\omega(1)$ and less than $d$, to get a family of $(d, 2^{o(d)}, k-1, k)$ embeddings.
         Again picking $d=\omega(\log n)$ small enough in OVP, we have by lemma~\ref{lem:emb+join} that $c\le (k-1)/k=1-1/k=1-o(1)$ is hard.
         Notice that this means any $c$ not bounded away from $1$ is hard.
   \end{enumerate}
\end{proof}

For the bounds parametrized by $\log(s)/\log(cs)$, we need to tweak our families slightly differently.
This in turn allows for hardness for shorter vectors than used in the previous results.

\begin{proof} (Theorem~\ref{thm:nojoin2})

   For embedding 1 the result follows directly as $\log(s/d)/\log(cs/d)\to0$ as $c\to0$.

   We have to remember that the results in theorem~\ref{thm:nojoin} are stated in terms of normalized $s$.
   For embedding 2 we calculate:
   \begin{align*}
      \frac{\log(s/d_2)}{\log(cs/d_2)}
      &= \frac{q\log(2/9) + q/\sqrt{d} - \log2}{q\log(2/9)} \\
      &= 1 - \frac{1}{\log(9/2)\sqrt{d}} + \frac{\log2}{q\log(9/2)} \\
      &= 1 - o\left(1/\sqrt{\log n}\right)
   \end{align*}
   Where in the last step we have taken $q=\sqrt{d}$ and $d=\omega(\log n)$ as by the OVP conjecture.

   It is important to notice that we could have taken $q$ much larger, and still satisfied lemma~\ref{lem:emb+join}.
   However that wouldn't have improved the result, except by more quickly vanishing second order asymptitic terms.
 What we instead gain from having $d_2=(9d)^{\sqrt{d}}$ is that, as one can verify going through the lemma, we show hardness for any join algorithm running in time $d^{o(\frac{\log{d}}{\log\log^2 d})}n^{1+\alpha-\epsilon}$.
 That is, the hardness holds even for algorithms with a much higher dependency on the dimension than polynomial.

   Similarly, we calculate for embedding 3:
\begin{align*}
 \frac{\log(s/d_2)}{\log(cs/d_2)}
 &= \frac{\log\frac{k}{k2^{d/k}}}{\log\frac{k-1}{k2^{d/k}}} \\
 &= 1 - \frac{k\log(1+1/(k-1))}{d+k\log(1+1/(k-1))} \\
 &= 1 - 1/d + O(1/(kd)) \\
 &= 1 - o(1/\log n)
 \end{align*}
 Where we have taken $k=d$ and $d=\omega(\log n)$ as by the OVP conjecture.

 Taking $k=d$ means that $d_2$ is only $2d$.

\end{proof}

\section{Limitations of LSH for IPS}

 We provide an upper bound on the gap between $P_1$ and $P_2$ for an $(s,cs,P_1,P_2)$-asymmetric LSH for signed/unsigned IPS.  For the sake of simplicity we assume the  data and query domains to  be the $d$-dimensional balls of radius 1 and $U\geq 1$, respectively. 
The bound holds for a fixed set of data vectors, so it  applies also to data dependent LSH~\cite{andoni2015optimal}.
A consequence of our result is that there cannot exist an asymmetric LSH for any dimension $d\geq 1$ when the set of query vectors is unbounded, getting a result similar to that of~\cite{NeyshaburS15}, which however requires even the data space to be unbounded and $d\geq 2$.

We firsts show in Lemma~\ref{lm:p1p2} that the gap $P_1-P_2$ can be expressed as a function of the length $h$ of two sequences of query and data vectors with suitable collision properties. 
Then we provide the proof of the aforementioned Theorem~\ref{th:gap}, where we derive some of such sequences and then apply the lemma.

\begin{figure*}
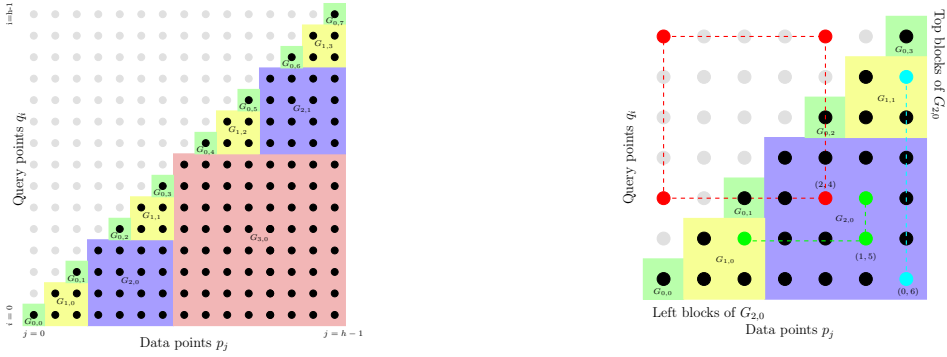

\begin{subfigure}{0.45\textwidth}
\centering
{\scalebox{0.45}{\input{grid_import.tex}}}
\end{subfigure}
\begin{subfigure}{0.45\textwidth}
\centering
{\scalebox{0.45}{\input{grid2_scaled_import.tex}}}
\end{subfigure}
\caption{\label{fig:matrixpartitioning} On the left, a $15\times 15$ grid: black nodes are $P_1$-nodes, gray nodes are $P_2$-nodes; the colored blocks denote the partitioning of the lower triangle into squares. On the right, a zoom of the $G_{2,0}$ square and of its left and top squares: the red nodes collide under a $(2,4)$-shared function; the green nodes collide under a $(1,5)$-partially shared function; the cyan node collide under a $(0,6)$-proper function (specifically, row proper).}
\end{figure*}

\begin{lemma}\label{lm:p1p2}
Suppose that there exists a sequence of data vectors $\P=\{p_0,\ldots, p_{n-1}\}$ and a sequence of query vectors $\Q=\{q_0,\ldots, q_{n-1}\}$ such that $q_i^Tp_j \geq s$ if $j\geq i$ and $q_i^Tp_j \leq cs$ otherwise (resp., 
$|q_i^Tp_j| \geq s$ if $j\geq i$ and $|q_i^Tp_j| \leq cs$ otherwise)
. Then any $(s,cs,P_1, P_2)$-asymmetric LSH for signed IPS (resp., unsigned IPS) must satisfy $P_1-P_2\leq 1/(8 \log n)$.
\end{lemma}
\begin{proof}
For the sake of simplicity we assume that $n=2^\ell -1$ for some $\ell\geq 1$; the assumption can be removed  by introducing floor and ceiling operations in the proof. 
Let $\H$ denote an $(s, cs, P_1, P_2)$-asymmetric LSH  family of hash functions, and let $h$ be a function in $\H$. The following argument works for signed and unsigned IPS.

Consider the $n\times n$ grid representing the collisions between $\Q\times \P$, that is, a node  $(i,j)$ denotes the query-data vectors $q_i$ and $p_j$. 
We say that a node $(i,j)$, with $0\leq i,j< n$, collides under $h$ if vectors $q_i$ and  $p_j$ collide under $h$. 
By definition of asymmetric LSH, all nodes with $j\geq i$ must collide with probability at least $P_1$, while the remaining nodes collide with probability at most $P_2$.
We use  \emph{lower triangle} to refer to the part of the grid with $j\geq i$ and \emph{$P_1$-nodes} to refer to the nodes within it; we refer to the remaining nodes as  \emph{$P_2$-nodes}.

We partition the lower triangle into squares of exponentially increasing side as shown in Figure~\ref{fig:matrixpartitioning}.
Specifically, we split the lower triangle into \emph{squares  $G_{r,s}$} for every $r$ and $s$ with
$0\leq r < \log (n+1)=\ell \text{ and } 0\leq s < (n+1)/2^{r+1}=2^{\ell-r-1},$ 
where $G_{r,s}$ includes all nodes in the square of side $2^r$ and top-left node $((2s+1)2^r-1, (2s+1)2^r-1)$. 
For a given square $G_{r,s}$, we define the \emph{left squares} (resp., \emph{top squares}) to be the set of squares that are on the left (resp., top) of  $G_{r,s}$.
We note that the left squares (resp.,  top squares) contain $2^{r-i-1}$ squares of side $2^i$ for any $0\leq i < r$ and all $P_1$-nodes with 
$s 2^{r+1}\leq i,j < (2s+1)2^r-1$
(resp.,  $(2s+1)2^r-1< i,j \leq (s+1)2^{r+1}-2$) .

We define the \textit{mass} $m_{i,j}$ of a node $(i,j)$ to be the collision probability, under $\mathcal H$, of  $q_i$ and $p_j$. 
We split the mass of a $P_1$-node into three contributions called shared mass, partially shared mass, and proper mass, all defined below.
Consider each $P_1$-node $(i,j$) and each function $h\in \H$ where $(i,j)$ collides. 
Let $G_{r,s}$ be the square containing $(i,j)$ and 
let $K_{h,i,j}$ denote the set of $P_1$-nodes $(i',j')$ on the left side of the same row or on the top of the same column of $(i,j)$ (i.e., $i'=i$ and $i\leq j'< j$, or $j'=j$ and $i < i' \leq j$) and with the same hash value of $(i,j)$ under $h$ (i.e.,  $h(i)=h(j)=h(i')=h(j')$). 
Clearly all nodes in $K_{h,i,j}$ collide under $h$. 
For the given node $(i,j)$, we classify $h$ as follows (see Figure~\ref{fig:matrixpartitioning} for an example):
\begin{itemize}
\item \emph{$(i,j)$-shared function.} $K_{h,i,j}$ contains at least a node $(i,j')$ in a left square, and at least a node $(i',j)$ in a top square. 

\item \emph{$(i,j)$-partially shared function.} Function $h$ is not in case 1 and $K_{h,i,j}$ contains at least a node node $(i,j')$ with $j'<j$, and at least a node $(i',j)$ with $i'>i$. 
That is, $K_{h,i,j}$ contains only nodes in $G_{r,s}$ and in the left blocks, or only nodes in $G_{r,s}$ and in the top blocks.

\item \emph{$(i,j)$-proper function.} $K_{h,i,j}$ contains no points $(i,j')$ for any $i\leq j'<j$  or contains no points $(i',j)$ for any $i< i'\leq j$. 
That is, $K_{h,i,j}$  cannot contain at the same time a point in a left square and a point in a top square.
Function $h$ is said row (resp., column) proper if there are no nodes in the same row (resp., column). 
We break ties arbitrary but consistently if $K_{h,i,j}$ is empty.  
\end{itemize}
The  \textit{shared mass} $m^{s}_{i,j}$ is the sum of probabilities of all $(i,j)$-shared functions.
The \textit{partially shared mass} $m^{ps}_{i,j}$ is the sum of probabilities of all $(i,j)$-partially shared functions.
The \textit{proper mass} $m^{p}_{i,j}$ is the sum of probabilities of all  $(i,j)$-proper functions (the row/column proper mass includes only row/column proper functions).
We have  $m_{i,j}= m^{p}_{i,j}+m^{ps}_{i,j}+m^s_{i,j}$.
The \emph{mass} $M_{r,s}$ of a square $G_{r,s}$ is the sum of the masses of all its nodes, while the \emph{proper mass} $M^p_{r,s}$  is the sum of  proper masses of all its nodes.
The sum of  row proper masses of all nodes in a row  is at most one since a function $h$ is row proper for at most one node in a row.
Similarly, the sum of column proper masses of all nodes in a column is at most one. 
Therefore, we  have that $\sum_{r,s} M^p_{r,s}\leq 2n$.

We now show that  $\sum_{(i,j)\in G_{r,s}} m^s_{i,j} \leq 2^{2r} P_2$ for every $G_{r,s}$.
Consider a node $(i,j)$ in a given $G_{r,s}$.
For each $(i,j)$-shared function $h$ there is a $P_2$-node colliding under $h$: indeed, $K_{h,i,j}$ contains nodes $(i,j')$ in the left blocks and $(i',j)$ in the top blocks with $h(i)=h(j)=h(i')=h(j')$ (i.e., $s 2^{r+1}\leq j' < (2s+1)2^r-1$ and $(2s+1)2^r-1< i' \leq (s+1)2^{r+1}-2$); then node $(i',j')$ is a $P_2$-node since $i'>j'$ and collides  under $h$.
By considering all nodes in $G_{r,s}$, we get that all the $P_2$-nodes that collide in a shared function are in the square of side $2^{r-1}$ and bottom-right node in $((2s+1)2^r, (2s+1)2^r-2)$.
Since these $P_2$-nodes have total mass at most $2^{2r}P_2$, the claim follows.

We now prove that $\sum_{(i,j)\in G_{r,s}} m^{ps}_{i,j} \leq  2^{r+1} M^p_{r,s}$. 
A  $(i,j)$-partially shared function is $(i',j)$ or $(i,j')$-proper  for some $i'<i$ and $j'>j$, otherwise there would be a node in  left blocks and a node in top blocks that collide with $(i,j)$ under $h$, implying that $h$ cannot be partially shared.
Since an $(i,j)$-proper function is partially shared for at most $2^{r+1}$ nodes in  $G_{r,s}$, we get 
$$\sum_{(i,j)\in G_{r,s}} m^{ps}_{i,j} \leq 2^{r+1} \sum_{(i,j)\in G_{r,s}} m^{p}_{i,j} = 2^{r+1} M^p_{r,s}.$$
By the above two bounds, we get
$$
M_{r,s} \leq \sum_{(i,j)\in G_{r,s}} \hspace{-.6em} m_{i,j}^p+m_{i,j}^{ps}+m_{i,j}^s \leq (2^{r+1}+1) M^p_{r,s}  + 2^{2r} P_2.
$$ 
Since $M_{r,s}\geq 2^{2r}P_1$ we get $M^p_{r,s} \geq (2^{r-1}-1) (P_1 - P_2)$.
By summing among all squares, we get
$$2n \geq \sum_{r=0}^{\ell-1} \sum_{s=0}^{2^{\ell-r-1}-1} M^p_{r,s} >  (P_1 - P_2)\frac{ n \log n}{4}$$ 
from which  the claim follows.
\end{proof}
We are now ready to prove Theorem~\ref{th:gap}.
\begin{proof} (Theorem~\ref{th:gap})
The upper bounds to $P_1-P_2$ in the different cases follow by applying Lemma~\ref{lm:p1p2} with different sequences of query and data vectors.  
We anticipate that in all three cases the gap $P_1-P_2$  becomes $0$ if the query ball is unbounded (i.e., $U=+\infty$), and hence there cannot exist an asymmetric LSH with $P_1>P_2$.

\emph{First case.} We now show that there exist data and query sequences of length $n=\BT{m d}$, with $m=\BT{\log_{1/c}(U/s)}$, for signed and unsigned IPS in $d\geq 1$ dimensions if $s=\BO{U/\sqrt{d}}$. Note that $m\geq 1$ since we assume $s\leq cU$.
As a warm-up, we start with $d=1$. Let $Q=\{q_i, \forall \;  0\leq i < n\}$ and $P=\{p_j, \forall \;  0\leq j < n\}$ with  
\begin{align}\label{eq:monodim}
q_i = Uc^i, \qquad 
p_j= {s}/({Uc^j}).
\end{align} 
Let $p_j\in P$ and $q_i\in Q$.
We get $p^T_j q_i = c^{i-j} s$: if $j\geq i$ then  $p^T_j q_i\geq s$ and  $p^T_j q_i\leq cs$ otherwise. Data and query vectors are respectively contained in the unit ball and in the ball of radius $U$ since $i,j< \BT{\log_{1/c}(U/s)}$. Being the sequences $P$ and $Q$ of length $m$, the claim follows.

Let now $d\geq 2$ and assume for the sake of simplicity  $d=2d'$ (the general case just requires some more tedious computations). 
Consider the following sequences $Q_k=\{q_{i,k}, \forall\; 0\leq i < m\}$ and  $P_k=\{p_{j,k}, \forall\; 0\leq j < m\}$ for each $0\leq k <d'$, where $q_{i,k}$   and $p_{j,k}$ are $d$-dimensional vectors defined as follows.
Denote with $q_{i,k}[t]$ the $t$-th coordinate of $q_{i,k}$, for $0\leq t < d$ (similarly for $p_{j,k}$).
For vector $q_{i,k}$ we have: $q_{i,k}[2k]=Uc^i$; $q_{i,k}[2t+1]=2s$ for each $k \leq t < d'$;  remaining positions are set to 0.
For vector $p_{i,k}$ we have: $p_{i,k}[2k]=s/(Uc^i)$;  $p_{i,k}[2k-1]=1/2$ (only when  $k> 0$);  remaining positions are set to 0. 
Intuitively, these data and query sequences follow by constructing the 1-dimensional sequences in Equation~\ref{eq:monodim} on $d'$ orthogonal dimensions and then by suitably translating each sequence.
 As an example, for $d=6$ we get:
\begin{align*}
&q_{i,0}=(Uc^i,2s,0,2{s},0,2s)& & p_{j,0}=(s/(Uc^j),0,0,0,0,0);\\
&q_{i,1}=(0,0,Uc^i,2{s},0,2s)& & p_{j,1}=(0, 1/2,s/(Uc^j),0,0,0);\\
&q_{i,2}=(0,0,0,0,Uc^i,2{s})& & p_{j,2}=(0,0,0,1/2,s/(Uc^j),0).
\end{align*}
The query and data sequences $Q=\{Q_0, \ldots Q_{d'-1}\}$ and $P=\{P_0, \ldots P_{d'-1}\}$ satisfy the hypothesis of Lemma~\ref{lm:p1p2}.
Indeed, it can be verified that: $p^T_{j,\ell}q_{i,\ell}=sc^{i-j}$ and thus $p^T_{j,\ell}q_{i,\ell}\geq s$ if $j\geq i$ and $p^T_{j,\ell}q_{i,\ell}\leq cs$ otherwise;
 $p^T_{j,\ell'}q_{i,\ell}=0$ if $\ell'<\ell$; $p^T_{j,\ell'}q_{i,\ell}\geq s$ if $\ell'> \ell$.
Further, when $s\leq U/(2 \sqrt{d})$ data and query vectors are contained in balls with radius 1 and $U$ respectively, with the exception of vectors  $q_{i,k}$ for $0\leq i <  1/(2\log (1/c))$ and $0\leq k <d'$   which are contained in a ball of radius $2U$. However, these query  vectors and the respective data vectors can be removed from the above sequences without affecting the asymptotic length. We thus get two sequences of length $n=(m- 1/(2\log (1/c)))d'=\BT{d\log_{1/c}(U/s)}$, the claim follows.
Since all inner products are non negative, the upper bound on $P_1-P_2$ holds for signed and unsigned IPS.

\emph{Second case.}
Longer query and data sequences, with length $n=\BT{m d}$ for $m=\BT{\sqrt{U/(s(1-c))}}$, can be constructed for signed IPS when $d\geq 2$. .
We start considering the case $d=2$. Let $Q=\{q_i, \forall \;  0\leq i < m\}$ and $P=\{p_j, \forall \;  0\leq j < m\}$ with 
\begin{equation}\label{eq:monodim2}
\begin{aligned}
q_i&=\left(\sqrt{sU}(1-(1-c)i),\sqrt{sU(1-c)}\right),\\
p_j&=\left(\sqrt{\frac{s}{U}}, j\sqrt{\frac{s(1-c)}{U}}\right).
\end{aligned}
\end{equation}
We observe that these sequences are similar to the one used in~\cite{NeyshaburS15}.
We have  $p_j^T q_i=s(1-c)(j-i)+s$: then,
$p_j^T q_i\geq s$ if $j\geq i$ and $p_j^T q_i\leq cs$ otherwise. If $s\leq U/2$, data and query vectors are within  balls of radius respectively 1 and $U$.

Let now $d\geq 2$ and assume for the sake of simplicity  $d=2d'$ (the general case just requires some more tedious computations). 
Consider the following sequences $Q_k=\{q_{i,k}, \forall\; 0\leq i < m\}$ and  $P_k=\{p_{j,k}, \forall\; 0\leq j < m\}$ for each $0\leq k <d'$, where $q_{i,k}$   and $p_{j,k}$ are $d$-dimensional vectors defined as follows.
For vector $q_{i,k}$ we have:  $q_{i,k}[2k]=\sqrt{sU}(1-(1-c)i)$; $q_{i,k}[2k+1]=\sqrt{sU(1-c)}$; $q_{i,k}[2t]=\sqrt{Us}$ for each $k < t < d'$; remaining positions are set to 0.
For vector $p_{i,k}$ we have: $p_{i,k}[2k]=\sqrt{{s}/{U}}$; $p_{i,k}[2k]=j\sqrt{{s(1-c)}/{U}}$;  remaining positions are set to 0. 
We observe that the two sequences follow by constructing the 2-dimensional sequences in Equation~\ref{eq:monodim2} on $d'$ orthogonal planes and then suitably translate.
Then, it follows that data and query sequences $P=\{P_0, \ldots P_{d'-1}\}$ and $Q=\{Q_0, \ldots Q_{d'-1}\}$ satisfy the hypothesis of Lemma~\ref{lm:p1p2} and they are  respectively contained in balls or radius one and $U$ respectively if $s\leq U/(2d)$. Being $m=nd'=\BO{d\sqrt{U/(s(1-c))}}$ and the claim follows.
We observe that the above sequences may generate large negative inner products and then they cannot be used for unsigned IPS.

\emph{Third case.}
Finally, we provide an upper bound on $P_1-P_2$ for signed and unsigned IPS that holds for $d\geq \BT{U^5/(c^2s^5)}$ by providing  data and query sequences of length $n=2^{\sqrt{U/(8s)}}$. 
Suppose there exists a family $\mathcal Z$ of $2n-1$ vectors such that $| z_i^T z_j | \leq \epsilon$ and $(1-\epsilon) \leq z_i^T z_i \leq (1+\epsilon)$ for any $z_i\neq z_j$, for  $\epsilon=c/(2\log^2 n)$.
It can be shown with the Johnson-Lindenstrauss  lemma that such a family exists when $d=\BOM{\epsilon^{-2}\log n}$ (for an analysis see e.g.~\cite{PaghSSS15}).
For notational convenience, we denote the vectors in $\mathcal Z$ as follows: $z_{b_0}, z_{b_0,b_1}, \ldots, z_{b_0,b_1, \ldots,b_{\log n-1}}$ for each possible value $b_0,\ldots, b_{\log n-1}\in\{0,1\}$.
Let ${b}_{i,\ell}$ denote the $\ell$-th bit of the binary representation of $i$ and with $\bar{b}_{i,\ell}$ its negation, where we assume $\ell=0$ to be the most significant bit. 
Let $Q=\{q_i, \forall \;  0\leq i < n\}$ and $P=\{p_j, \forall \;  0\leq j < n\}$ with  
\begin{align*}
q_i = \sqrt{2sU} \sum_{\ell=0}^{\log n-1} \bar{b}_{i,\ell} z_{{b}_{i,0},\ldots {b}_{i,\ell-1}, \bar{b}_{i,\ell}}\\
p_j = \sqrt{2s/U} \sum_{\ell=0}^{\log n-1} b_{j,\ell} z_{{b}_{j,0},\ldots {b}_{j,\ell-1}, {b}_{j,\ell}}
\end{align*}
Since the inner product of two distinct vectors in $\mathcal Z$ is in the range $[-\epsilon,\epsilon]$, we have that $p_j^T q_i$ can be upper bounded as 
\begin{align*}
p_j^T q_i\leq &\epsilon {2s} (\log^2 n-\log n)+ \\&+ 2s\sum_{\ell=0}^{\log n-1} 
{b}_{j,\ell}\bar{b}_{i,\ell}  z_{{b}_{j,0},\ldots {b}_{j,\ell-1}, {b}_{j,\ell}}
 z_{{b}_{i,0},\ldots {b}_{i,\ell-1}, \bar{b}_{i,\ell}}
\end{align*}
Suppose $i>j$. Then there exists a bit position $\ell'$ such that ${b}_{i,\ell'}=1$, ${b}_{j,\ell'}=0$ and ${b}_{i,\ell}={b}_{j,\ell}$ for all $\ell<\ell'$. 
We get  ${b}_{j,\ell} \bar{b}_{i,\ell}=0$ for all $\ell\leq \ell'$ and  
$z_{{b}_{j,0},\ldots {b}_{j,\ell-1}, {b}_{j,\ell}}\neq z_{{b}_{i,0},\ldots {b}_{i,\ell}, \bar{b}_{i,\ell}}$ for all $\ell> \ell'$.
It then follows that  $p_j^T q_i < \epsilon {2s} \log^2 n$ when $i>j$.
On the other hand we get that $p_j^T q_i$ can  be lower bounded as
\begin{align*}
p_j^T q_i\geq& -\epsilon {2s} (\log^2 n-\log n)+ \\&+ 2s\sum_{\ell=0}^{\log n-1} 
{b}_{j,\ell}\bar{b}_{i,\ell}  z_{{b}_{j,0},\ldots {b}_{j,\ell-1}, {b}_{j,\ell}}
 z_{{b}_{i,0},\ldots {b}_{i,\ell-1}, \bar{b}_{i,\ell}}
\end{align*}
Suppose $i\leq j$. Then there exists  an index $\ell'$
such that ${b}_{j,\ell'}=1$, ${b}_{i,\ell'}=0$ and ${b}_{j,\ell}={b}_{i,\ell}$ for all $\ell<\ell'$. 
We get ${b}_{j,\ell'} \bar{b}_{i,\ell'}=1$
and $z_{{b}_{j,0},\ldots {b}_{j,\ell'-1},{b}_{j,\ell'}}= z_{{b}_{i,0},\ldots {b}_{i,\ell'-1}, \bar{b}_{i,\ell'}}$. 
It then follows that  $p_j^T q_i \geq -\epsilon  {2s}  \log^2 n + 2s$.
When $d=\BOM{(\log^5 n)/c^2}$, we can set $\epsilon=c/(2\log^2 n)$.
From the above lower and upper bounds, it then follows that   $p_j^T q_i\leq cs$ if $j<i$ and $p_j^T q_i\geq s$ if $j\geq i$ and hence the sequences $Q$ and $P$ satisfy the hypothesis of Lemmma~\ref{lm:p1p2}.
Finally, we observe that the data and query vectors in $P$ and $Q$ are respectively contained in balls of radius $1$ and $U$: each $q_i$ (resp., $p_j$) is given by the sum of at most $\log n$ vectors whose norm is not larger than $\sqrt{2sU}(1+\epsilon)$ (resp., $\sqrt{2s/U}(1+\epsilon)$); being  $n=2^{\sqrt{U/(8s)}}$ the claim follows.
\end{proof}

\section{Upper bounds}\label{sec:up}

This section contains three observations with implications for IPS join and its indexing version. We first notice in Section~\ref{up1} that by plugging the best known LSH for $\ell_2$ distance on a sphere~\cite{andoni2015optimal} into a reduction presented in~\cite{Bachrach14,NeyshaburS15}, we get a data structure based on LSH for signed MIPS with search time exponent
$ \rho = (1 - s)/(1 + (1 - 2c)s). $

Then, in Section~\ref{up2}, we show how to circumvent the results in~\cite{NeyshaburS15,ShrivastavaL14} showing that symmetric LSH is not possible when the data and query domains coincide (while an asymmetric LSH does exist).
We use an slightly modified definition of LSH that disregards the collision probability of 1 for pairs of identical vectors, and assume that vectors are represented with finite precision.
The LSH construction uses explicit incoherent matrices built using Reed-Solomon codes~\cite{NelsonNW12} to implement a symmetric version of the reduction in~\cite{Bachrach14,NeyshaburS15}.

Finally, in Section~\ref{up3} we solve unsigned $(cs,s)$ join using linear sketches for $\ell_p$ norms from~\cite{alexlpmanuscript}.
Given $\kappa \geq 2$ we obtain a approximation factor $c\geq 1/n^{1/\kappa}$ using $\BTO{dn^{2-2/\kappa}}$ time. 
Although this trade-off is not that strong, it is not far from the conditional lower bound in Theorem~\ref{thm:nojoin}.

\subsection{Asymmetric LSH for signed IPS}\label{up1}

\begin{figure*}[th]
    \centering
    \includegraphics[width=0.9\textwidth]{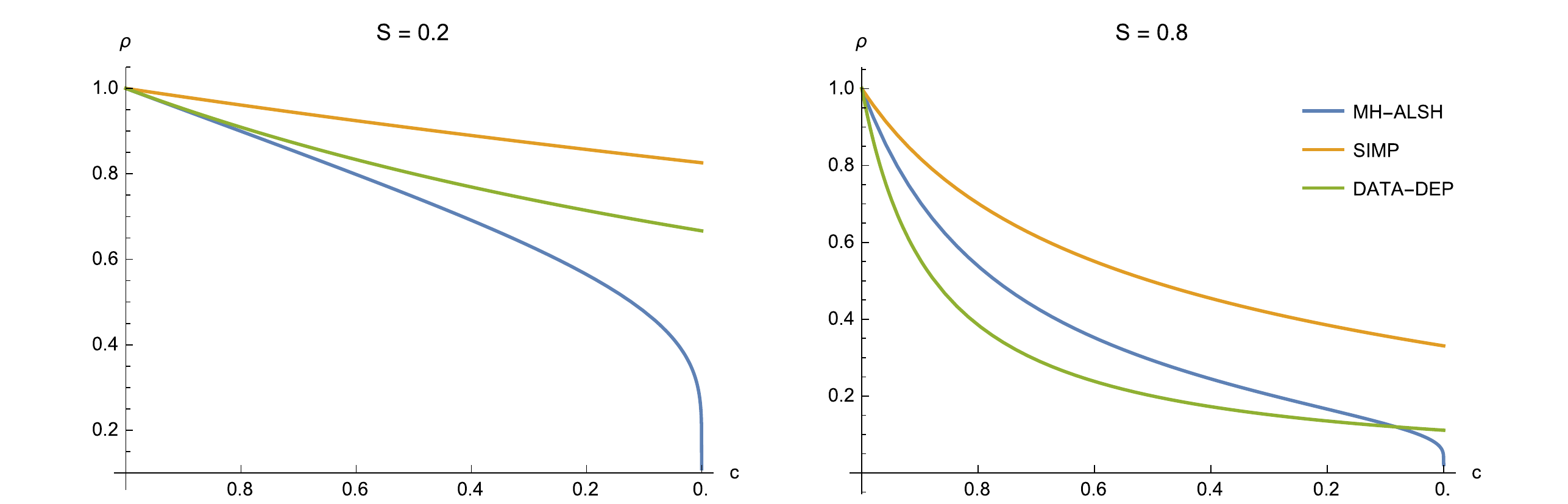}
    \caption{\label{fig:rho}Our $\rho$ value (DATA-DEP) compared to that of \cite{NeyshaburS15} (SIMP) and the binary data only of \cite{shrivastava2015asymmetric} (MH-ALSH).}
\end{figure*}

We assume the data and query domains to be $d$-dimensional balls with respective radius 1  and $U$. 
Vectors are embedded into a $(d+2)$-dimensional unit sphere using the asymmetric map as in~\cite{NeyshaburS15}: a data vector $p$ is mapped to $ (p,\sqrt{1-||p||^2},0)$, while a query $q$ is mapped to $ (q/U,0,\sqrt{1-||q||^2/U^2})$. 
This transformation  just scales the inner product by a factor $U$, and hence signed inner product search can be seen as an instance of ANN in $\ell_2$ with distance threshold $r=\sqrt{2(1 - s/U)}$ and approximation $c'=\sqrt{(1 - cs/U)/(1-s/U)}$. 
The latter can be solved in space $O(n^{1 + \rho} + dn)$ and query time $O(n^{\rho})$ using the LSH construction of~\cite{andoni2015optimal}. We  get the following $\rho$ value (for the LSH gap as well as for the exponent of the running time):  
\begin{equation}
\label{mips_rho}
        \rho = \frac{1}{2c'^2 -1} = \frac{1 - s/U}{1 + (1 - 2c)s/U} \enspace .
\end{equation}

In Figure~\ref{fig:rho}, we plot the $\rho$ values of three LSH constructions: the one proposed here (with $U=1$), the one from~\cite{NeyshaburS15}, and the one from~\cite{shrivastava2015asymmetric}. The latter
works only for binary vectors. We point out that our bound is always stronger than the one from~\cite{NeyshaburS15} and sometimes
stronger than the one from~\cite{shrivastava2015asymmetric}, despite that the latter is tailored for binary vectors.
The latter conclusion is somewhat surprising, since the data structure we obtain works for non-binary vectors as well.

We point out that in practice one may want to use a recent LSH family from~\cite{practicalballcarving}
that---{both in theory and in practice}---is superior to the hyperplane LSH from~\cite{Cha02} used in~\cite{NeyshaburS15}.  

\subsection{Symmetric LSH for almost all vectors}\label{up2}
Neyshabur and Srebro~\cite{NeyshaburS15} show that an asymmetric view on LSH for signed IPS is required.
Indeed they show that a symmetric LSH for signed IPS does not exist when  data and query domains are balls of the same radius, while an asymmetric LSH does exist.
(On the other hand, when the data domain is a ball of given radius $U$ and the query domain is a sphere of same radius, a symmetric LSH does exist.)
In this section we show that even when data and query spaces coincide a nontrivial symmetric LSH does exist if we disregard the trivial collision probability of 1 when data and query vectors are identical. 

We first show how to reduce signed IPS to the case where data and query vectors lie on a unit sphere.
The reduction is deterministic and maintains inner products up to an additive error $\eps$ for all vectors $x,y$ with $x\neq y$. 
We then plug in any Euclidean LSH for ANN on the sphere, for example the one from~\cite{andoni2015optimal}.
This reduction treats data and query vectors identically, unlike the one from~\cite{NeyshaburS15}, and thus we are able to obtain a symmetric LSH.

Assume that all the coordinates of all the data and queries are encoded as $k$-bit numbers and that the data and query vectors are in the unit ball.
The idea is the following. There are at most $N = 2^{O(dk)}$ possible data vectors and queries. 
Imagine a collection of $N$ unit vectors $v_1, \ldots, v_N$ such that for every $i \ne j$ one has $|v_i^T v_j| \leq \eps$.
Then, it is easy to check that a map of a vector $p$ to $f(p)= (p,  \sqrt{1 - \|p\|^2} \cdot v_p)$ maps a vector from a unit ball to a unit sphere and, moreover, for $p \ne q$ one has
$
|f(p)^T f(q)-p^T q|  \leq \eps.
$

What remains is to construct such a collection of vectors~$v_i$. 
Moreover, our collection of vectors must be explicit in a strong sense:
we should be able to compute $v_u$ given a vector $u$ (after interpreting it as an $dk$-bit string).
Such a constructions are well known, e.g., in~\cite{NelsonNW12} it is shown how to build such vectors using Reed-Solomon codes.
The resulting dimension is $\BO{ \eps^{-2}\log N} = \BO{kd/ \eps^2}$~\cite{LarsenN14,NelsonNW12}.

After performing such a reduction we can apply any state-of-the-art LSH (or data structure for ANN) for $\ell_2$ norm on a sphere, e.g.~from~\cite{andoni2015optimal,practicalballcarving}, with distance threshold $r^2=2(1-s+\epsilon)$, approximation factor $c'^2=(1-cs-\epsilon)/r^2$. If $\eps$ is sufficiently small we get a $\rho$ value close to the one in (\ref{mips_rho}).
The final result is therefore a symmetric LSH for symmetric domains that does not provide any collision bound for all pairs $(q,p)$ with $q=p$ since the guarantees on the inner product fail for these pairs.
This LSH can used for solving signed $(cs,s)$ IPS as a traditional LSH~\cite{AndoniI08}, although it is required an initial step that verifies whether a query vector is in the input set and, if this is the case, returns the vector $q$ itself if $q^Tq\geq s$.

\subsection{Unsigned IPS via linear sketches}
\label{up3}
In this section we propose a linear sketch for \emph{unsigned $c$-MIPS}, that can be used for solving unsigned $(cs,s)$ join. 
The {unsigned $c$-MIPS}  is defined as follows: given a set $\P\subset \mathbb{R}^d$ of $n$ vectors, construct a data structure that efficiently returns, for a given query vector $q$, a vector $p\in P$ where $|p^Tq| \geq c (p'^Tq)$, where $p'$ is the vector in $P$ with maximum absolute inner product with $q$.
The unsigned $(cs,s)$ join between sets $P$ and $Q$ can be computed by constructing a data structure for unsigned $c$-MIPS for vectors in $P$ and then performs a query for each vector in $Q$.

Of independent interest, we notice that unsigned $c$-MIPS can be solved by a data structure for unsigned $(cs,s)$ search. Let $\mathcal D$ be a data structure for unsigned $(cs,s)$ search on the data set $P$, and suppose we are given a query $q$ and the promise that there exists $p'\in P$ such that $p'^Tq>\gamma$.
Then,  unsigned $c$-MIPS can be solved by performing on $\mathcal D$ the queries $q/c^i$ for any $0\leq i \leq \lceil \log_{1/c} (s/\gamma)\rceil$. 
Intuitively, we are scaling up the query $q$ until the largest inner product becomes larger than the threshold $s$.
We notice that~$\gamma$ can  be also considered  as the smallest inner product that can be stored according to the numerical precision of the machine.

Our data structure requires  $\BTO{dn^{2-2/\kappa}}$ construction time and $\BTO{dn^{1-2/\kappa}}$ query time and provide a $c\geq 1/n^{1/\kappa}$ approximation with high probability, for any $\kappa \geq 2$. 
This gives an algorithm  for  unsigned $(cs,s)$ join  on two sets of size $n$ requiring time  $\BTO{dn^{2-2/\kappa}}$.
As shown in Theorem~\ref{thm:nojoin}, we are unlikely to significant improve further the approximation factor  if the OVP conjecture is true.

First, suppose we are only interested in approximating the value of $\max_p |q^t p|$ and not to find the corresponding vector.
Then, the problem is equivalent to estimating $\|Aq\|_{\infty}$, where $A$ is an $n \times d$ matrix, whose rows are data vectors.
This problem can be tackled using {linear sketches} (for an  overview see~\cite{Woodruff-sketchBook,AKR15}).
More specifically, we use the following result from~\cite{alexlpmanuscript}: for every $2 \leq \kappa \leq \infty$  there exists a distribution over $\widetilde{O}(n^{1 - 2 / \kappa}) \times n$ matrices $\Pi$ such that  for every $x \in \Rbb^n$ one has:
  $$
  \Pr_{\Pi} \left[(1-c)\|x\|_\kappa \leq \|\Pi x\|_{\infty}\leq (1+c)\|x\|_\kappa\right] \geq 0.99
  $$
for a suitable constant $0< c< 1$.  
Thus, to build a data structure for computing $\|Aq\|_{\infty}$, we sample a matrix $\Pi$ according to the aforementioned result in~\cite{alexlpmanuscript} and compute
the $\BTO{n^{1 - 2/\kappa}} \times d$ matrix $A_s = \Pi A$. 
Then, for every query~$q$, we compute $\|A_s q\|_{\infty}$ 
in time $\BTO{d \cdot n^{1 - 2/\kappa}}$, which is a $\BO{n^{1/\kappa}}$-approximation to $\|Aq\|_{\infty}$ with probability at least $0.99$.
Note that  we can reduce the probability
of error from $0.01$ to $\delta > 0$ as usual, by building $O(\log(1 / \delta))$ independent copies of the above data structure and reporting the median estimate.

We now consider the recovery of the vector that almost maximizes $|p^t q|$. We recover the index of the desired vector bit by bit. That is, for every bit index $0\leq i < \log n$, we consider every binary sequence $b$ of length $i$ and build a data structure for the dataset containing only the
vectors in $\mathcal{P}$ for which the binary representations of their indexes have prefix $b$. 
Although the number of data structures is $n$, the total required space is still $\BTO{dn^{1-2/\kappa}}$ since each vector appears in only $\log n$ data structures.
The claim stated at the beginning follows.

\section{Conclusion}
This paper has investigated different aspects of the complexity of \emph{approximate} similarity join with inner product.
In particular, we have related the hardness of this problem to the OVP conjecture.
Under some assumptions on $c$ and $s$, the proposed conditional lower bounds rule out algorithms for signed/unsigned $(cs,s)$ IPS join running in $n^{2-\epsilon}$ time, for a constant $\epsilon>0$, unless the OVP conjecture is false.
Nevertheless, the  data structures in section~\ref{sec:up} show that it still possible to reach weak subquadratic time, and even truly subquadratic time for small values of the approximation factor.

The hardness of signed/unsigned IPS holds even for weak approximation factors when the vector domain is $\{-1,1\}^d$.
Indeed, the result holds if $c\geq 0$ for  signed join, and if $c\geq e^{-o(\sqrt{\log n}/\log\log n)}$ for  unsigned join.
When $c<1/n^{\BOM{1}}$, the data structure for unsigned IPS in section~\ref{up3} reaches strongly subquadratic time and this gives evidence that the constraint on $c$ of the hardness result cannot be significantly relaxed.
On the other hand, when vectors are in the $\{0,1\}^d$ domain, a stronger assumption is required on the approximation factor.
In this case, the conditional lower bound holds for $c=1-o(1)$  and hence it does not rule out an algorithm running in $n^{2-\epsilon}$ time for a constant approximation factor.
We believe that a  different approach is required to show the hardness of IPS  for constant approximation: indeed, the proposed reduction from OVP to IPS in the $\{0,1\}^d$ domain strongly relies on the ability to distinguish inner products smaller than $k-1$ and larger than $k$ for some $k=\omega(1)$, implying $c\geq  1-o(1)$.
From an upper bound point of view, the LSH proposed in section~\ref{up1} improves upon the state of the art~\cite{shrivastava2015asymmetric} based on minwise hashing for different values (e.g., when $s\geq d/3$ and $c\geq 0.83$); however, it still gives weak subquadratic algorithm for signed/unsigned IPS  with constant $c$.
An interesting open question is therefore to assess if strongly subquadratic time is possible when the approximation is constant (or smaller) in the $\{0,1\}^d$ domain.

\bibliographystyle{abbrv}
\bibliography{paper}

\end{document}